\documentclass{aa} % for a referee version
\usepackage{graphicx}
\usepackage{txfonts}
\usepackage{bm}
\usepackage[colorlinks=true,linkcolor=blue,citecolor=blue]{hyperref}
\begin{document} 

\newcommand{\psj}{PSJ}
  
  \title{Dynamical tides in Jupiter and the role of interior structure}

   \author{Yufeng Lin
         % \inst{1}
          % \and
          %C. Ptolemy\inst{2}\fnmsep\thanks{Just to show the usage
          %of the elements in the author field}
          }

   \institute{Department of Earth and Space Sciences, Southern University of Science and Technology, Shenzhen 518055, China\\
              \email{linyf@sustech.edu.cn}
             }
\authorrunning{Lin}
%\titlerunning{Dynamical tides in Jupiter}

% \abstract{}{}{}{}{} 
% 5 {} token are mandatory
 
  \abstract
   { The Juno spacecraft has obtained highly accurate tidal Love numbers, which provide important constraints on the tidal response and interior structure of Jupiter.}
   {In order to exploit these observations, it is necessary to develop an approach to accurately  calculate the tidal response of Jupiter for a given interior model and to investigate the role of the interior structure.}
   {We directly solve  the linearized tidal equations of a compressible, self-gravitating, rotating, and viscous fluid body using a pseudo-spectral method. The Coriolis force is fully taken into account, but the centrifugal effect is neglected. We are able to simultaneously obtain the real and imaginary parts of the tidal Love numbers for a given planetary interior model.} 
   {{We calculated the tidal responses for three simplified interior models of Jupiter which may contain a compact rigid core or an extended dilute core. All of the models we consider can explain the fractional correction $\Delta k_{22}\approx -4\%$ due to dynamical tides, but they all have difficulties reconciling  the observed $\Delta k_{42}\approx -11\%$ for the high-degree tidal Love number. We show that the Coriolis force significantly modifies gravity modes in an extended dilute core at the tidal frequency relevant to the Galilean satellites. We demonstrate that the existence of a thin stable layer in the outer region would also influence the tidal responses of Jupiter.}}
   {}


   \maketitle
%
%-------------------------------------------------------------------
\section{Introduction}\label{sec:Intro}
Tidal interactions between Jupiter and the Galilean satellites play an important role in the orbital evolution of the system and the internal dynamics of the moons \citep{Lainey2009}. The highly active volcanic eruptions on Io are believed to be due to strong tides raised by Jupiter \citep{Peale1979}.  Meanwhile, tides are also raised in Jupiter by its moons, which are probably dominated by Io \citep{Gavrilov1977Icar}. The tidal response of a gaseous body such as Jupiter is conventionally treated as a hydrostatic deformation, which acquires a small phase lag  with respect to the tidal forcing due to dissipative processes. This is known as the equilibrium tide. However, the equilibrium tide alone does not suffice to account for the observed strong tidal dissipation in Jupiter \citep{Lainey2009}  and the gravitational perturbations recently measured by the Juno spacecraft \citep{Durante2020}.  

In fact, the equilibrium tide does not satisfy the momentum equation of tidal flows and thus corrections have to be made to fully account for the tidal response of Jupiter. The corrections to the equilibrium tide are collectively referred to as the dynamical tide, which usually involves wave-like motions in the planet  and depends on the tidal frequency as well as the interior structure \citep{Ogilvie2014}. The dynamical tide may provide extra channels of tidal dissipation and  produce additional gravitational perturbations in addition to the hydrostatic deformation. The Juno spacecraft has obtained highly accurate tidal Love numbers, $k_{lm}$ \citep{Durante2020}, which quantitatively characterize the tidal response of Jupiter to a tidal forcing component represented in spherical harmonics of degree $l$ and order $m$. The observed tidal Love numbers by Juno exhibit non-negligible discrepancies with respect to the theoretically calculated hydrostatic values \citep{Wahl2020ApJ}, suggesting that the dynamical tide has to be considered to explain the observed tidal response. Specifically, Juno observations found $\Delta k_{22}\approx -4\%$ for the dominant tidal component $l=2$ and $m=2$ and $\Delta k_{42}\approx -11\%$ for the high-degree tidal component  $l=4$ and $m=2$, where  $\Delta k_{lm}=(k_{lm}-k_{lm}^{(hs)})/k_{lm}^{(hs)}$  represents the fractional correction to the hydrostatic value $k_{lm}^{(hs)}$ \citep{Wahl2020ApJ,Idini2021PSJ,Idini2022PSJLost}. 
 
 As the dynamical tides are sensitive to the tidal frequency and the interior structure, the detected gravitational signatures of dynamical tides may provide important constraints on Jupiter's interior \citep{Idini2021PSJ,Lai2021PSJ,Idini2021PSJ,Idini2022PSJ,Dewberry2022ApJ}. 
Recent studies \citep{Idini2021PSJ,Lai2021PSJ} have revealed that  the discrepancy in $k_{22}$ can be mainly attributed to the Coriolis effect on the fundamental modes ($f$-modes). More recently, \cite{Idini2022PSJ} proposed that the resonant locking with a gravity mode in an extended dilute core can explain $\Delta k_{42}\approx -11\%$. This provides an independent constraint on the existence of a dilute core in Jupiter, which has also been suggested by the Juno measurements of gravitational moments of Jupiter \citep{Wahl2017GeoRL,Militzer2022PSJ}. However, the tidal constraint on the existence of a dilute core is uncertain. The calculation of the tidal response in \cite{Idini2022PSJ} inadequately treated the rotational (Coriolis) effect, which plays an important role in Jupiter's tidal responses because the tidal frequencies of Galilean satellites are comparable to the spin frequency of Jupiter. Including the Coriolis force introduces inertial waves in the neutrally buoyant regions \citep{Ogilvie2004ApJ,Wu2005ApJI} as well as mixed gravity waves and inertial waves (gravito-inertial waves) in the stably stratified region \citep{Dintrans1999,Xu2017PhRvD}.  The mechanism proposed by  \cite{Idini2022PSJ} is also struggling to reconcile both the real part (relevant to the gravitational perturbation)  and imaginary part (relevant to the tidal dissipation) of the tidal Love numbers.     

For this study, we developed a method to directly calculate the tidal response of a fully compressible, self-gravitating, rotating, and viscous fluid body. The Coriolis force is fully taken into account, but the centrifugal force is neglected, which allows us to numerically solve the problem in spherical geometry using a pseudo-spectral method based on spherical harmonic expansions \citep{Ogilvie2004ApJ,Lin2017MNRAS}. As we directly solve the tidally forced problem with explicit viscosity, we could  simultaneously obtain the real and imaginary parts of the tidal Love number for a given planetary interior model.  Our approach is different from recent studies on dynamical tides of Jupiter \citep{Lai2021PSJ,Idini2022PSJ,Dewberry2022ApJ}. They obtained the eigen modes of the inviscid fluid body first and then calculated the tidal Love number (only the real part) through projecting the tidal force onto each eigen modes. 
We consider three nominal interior models of Jupiter to investigate the dependence of the tidal response on the tidal frequency and the interior structure. 
{We focus on the effect of a compact rigid core, an extended dilute core, and a thin stably stratified layer in the outer region on tidal responses. All of the simplified models can explain the observed $\Delta k_{22}\approx-4\%$ as previous studies have shown. However, these simplified models cannot account for the observed $\Delta k_{42}\approx-11\%$. Resonances with gravito-inertial modes in an extended dilute core near the tidal frequency of Io can produce non-negligible dynamical correction to $ k_{42}$, but it is insufficient to explain the Juno observation based on our simplified model.}   

\section{Tidal model}
We consider linear tidal responses of a rotating  gaseous planet to a tidal potential component of $\Psi_l^m=\mathcal A (r/R)^l Y_l^m(\theta,\phi) \mathrm{e}^{-\mathrm{i}\omega t}$, where $\mathcal A$ is the tidal amplitude, $R$ is the radius of the planet, $Y_l^m(\theta,\phi)$ represents spherical harmonics, and $\omega$ is the tidal frequency. The resulting tides  of the planet produce an external gravitational potential perturbation $\Phi'=\mathcal B (R/r)^{l+1} Y_l^m(\theta,\phi) \mathrm{e}^{-\mathrm{i}\omega t}$ (and probably other spherical harmonic components). The ratio $K_l^m(\omega)=\mathcal B/ \mathcal A$ defines the tidal Love number, which depends on the tidal frequency. The tidal Love number $K_l^m$ is a complex number because there exists a phase lag between the forcing and the gravitational perturbations due to dissipative processes \citep{Ogilvie2014}. While the real part $k_{lm}=\mathrm{Re}[K_l^m]$ measures the in-phase gravitational perturbations with the tidal forcing, the imaginary part $\mathrm{Im}[K_l^m]$ quantifies the out-of-phase tidal response and is related to the dissipation rate. The ratio between the real and imaginary parts is related to the tidal quality factor
\begin{equation}\label{eq:Q}
Q=\mathrm{sgn}(\omega) \frac{k_{lm}}{\mathrm{Im}[K_l^m]} ,
\end{equation}
where  $\mathrm{sgn}(\omega)=\pm1$ is the sign function. Because the phase lag is  generally very small, that is $Q\gg 1$, the magnitude of the imaginary part is typically much smaller than the real part. 
For this study, we develope an approach to directly and simultaneously calculate the real and imaginary parts of the tidal Love number for a given planetary model. 

\subsection{Linearized equations} \label{sec:eq}
For a compressible, self-gravitating, and rotating fluid body which may contain a rigid core of radius $R_i$, linear perturbations to a tidal potential $\Psi \propto \mathrm{e}^{-\mathrm{i}\omega t}$ in the rotating frame are described by the following equations \citep[e.g.,][]{Ogilvie2004ApJ}:

\begin{equation} \label{eq:MomEq}
 -i \omega \bm u' =  -2\bm{\Omega\times u'}-\frac{1}{\rho_0}\bm\nabla P'+\frac{\rho'}{\rho_0^2}\bm \nabla P_0 -\bm \nabla \Phi' -\bm \nabla \Psi+\bm f_\nu,
\end{equation}
\begin{equation}
-i \omega \rho'+\bm{\nabla\cdot}(\rho_0\bm u')=0,
\end{equation}

\begin{equation}\label{adiabatic_eq}
-i \omega\left(\frac{P'}{\Gamma P_0}-\frac{\rho'}{\rho_0}\right)+\bm u'\cdot \left(\frac{1}{\Gamma}\nabla \ln P_0-\nabla {\ln \rho_0}\right)=0
\end{equation}
\begin{equation}\label{Possion_eq}
\nabla ^2 \Phi'=4\pi G \rho',
\end{equation}
where $\bm u$ is the velocity, $\bm \Omega$ the rotation rate, $\rho$ the density, $P$ the pressure, $\Gamma$ the adiabatic index, and $G$ the gravitational constant. In the above equations, the subscript $_0$ denotes physical quantities in the hydrostatic state (without tidal potential)  and the notations with the prime represent Eulerian perturbations induced by the tidal forcing. In the momentum equation (\ref{eq:MomEq}), we have explicitly included a viscous force $\bm f_\nu$  defined as 
\begin{equation}
\bm f_\nu=\frac{1}{\rho_0}\bm \nabla \cdot (2\mu \bm S),
\end{equation}
where $\mu$ is the dynamic shear viscosity (we neglected the bulk viscosity) and $\bm S$ is the strain-rate tensor:
\begin{equation}
\bm S= \frac{1}{2} \left[ \bm \nabla u'+(\bm \nabla \bm u')^T \right]-\frac{1}{3}(\bm \nabla \cdot \bm u')\mathbf{I}.
\end{equation}
We have included the viscous force in the momentum equation, but we neglected the viscous heating in the energy equation, that is the density and pressure perturbations were treated as adiabatic.  

For this study, we have fully taken the Coriolis force into account, but neglected the centrifugal distortion for numerical convenience. The centrifugal effect can be measured by $\epsilon=\Omega/\omega_{dyn}$, that is the ratio  between the spin frequency $\Omega$ and the dynamical frequency  $\omega_{dyn}=(GM/R^3)^{1/2}$. This ratio is not particularly small for Jupiter ($\epsilon=0.288$). Indeed, the centrifugal distortion of Jupiter has non-negligible contributions to the total Love number $k_{lm}$, especially for the high-degree Love number $k_{42}$ because the tidal response at $l=m=2$ can produce a gravitational perturbation at $l=4$ and $m=2$ in an oblate figure  \citep{Idini2022PSJLost}.  For the hydrostatic $k_{42}^{(hs)}$ of Jupiter due to Io, 93\% of the total value  is actually contributed by the centrifugal coupling with $k_{22}$ and only the remaining 7\% is produced by the tidal forcing at $l=4$ and $m=2$ \citep{Wahl2020ApJ,Idini2022PSJLost}. In this paper we do not aim to directly fit the  $k_{lm}$ observed by Juno, but rather focus on the fractional corrections $\Delta k_{lm}$ by the dynamical tides. In terms of the fractional correction $\Delta k_{lm}$, the centrifugal contribution to  $\Delta k_{22}$ can be neglected in leading order \citep{Lai2021PSJ}. However, the centrifugal contribution to $\Delta k_{42}$ cannot be neglected even in leading order because $k_{42}^{(hs)}$ is mostly contributed to by the centrifugal coupling with  $k_{22}$. This complicates the comparison between the calculated $\Delta k_{42}$ in a spherical figure and the observation. Nevertheless, the calculated  $\Delta k_{42}$ in a spherical figure can be multiplied by the factor 0.07 to account for Jupiter’s centrifugal coupling effect for qualitative comparisons with the observation \citep{Idini2022PSJ}. Such a comparison would assume that the tidally excited internal modes are not significantly modified by the centrifugal deformation.
 
By neglecting the centrifugal deformation, the unperturbed basic state is spherically symmetric, that is to say it depends on the radius $r$ only. Given the density $\rho_0(r)$ and pressure $P_0(r)$ profiles of the unperturbed state, the radial gravitational acceleration (inward) $g(r)$ and the Brunt-V\"ais\"al\"a frequency $N(r)$ are then determined by  
\begin{equation}
    g=\frac{d \Phi_0}{ d r} = -\frac{1}{\rho_0}\frac{d P_0}{dr},
\end{equation}
\begin{equation}
    N^2=g \left(\frac{1}{\Gamma}\frac{d \ln P_0}{dr}-\frac{d \ln \rho_0}{dr}\right).
\end{equation}

\subsection{Numerical method}
In order to obtain the complex Love numbers, we numerically solved Eqs. (\ref{eq:MomEq}-\ref{Possion_eq}) using a pseudo-spectral method  for the prescribed basic states, which are subject to the relevant boundary conditions. The numerical scheme is based on the method used in  previous studies \citep{Ogilvie2004ApJ,Lin2017MNRAS}, but we extended the method to solve the full set of linearized equations (\ref{eq:MomEq}-\ref{Possion_eq}) without making a low-frequency approximation \citep{Ogilvie2013MNRAS}. 
By  introducing  $h'=P'/\rho_0$  and eliminating the density perturbation $\rho'$, Eqs. (\ref{eq:MomEq}-\ref{Possion_eq}) can be reduced to the following equations:

\begin{eqnarray}\label{eq:u}
 -i \omega \rho_0\bm u' &= &-2\rho_0 \bm{\Omega\times u'}-\bm\nabla (\rho_0 h')+\bm{g}\nabla^2 \varphi'/(4 \pi G) \nonumber  \\
 & &-\rho_0 \bm \nabla \varphi'-\bm \nabla \Psi+\nabla\cdot (2 \mu \bm S),
\end{eqnarray}

\begin{equation} \label{eq:h}
  -i \omega  h'=-c_s^2 (N^2 u'_r/g+\bm{\nabla\cdot}(\rho_0\bm u')/\rho_0),
\end{equation}

\begin{equation} \label{eq:phi}
  -i \omega \nabla^2 \varphi'=-4 \pi G  \bm{\nabla\cdot}(\rho_0\bm u'),
\end{equation}
where $u_r'$ is the radial velocity perturbation and  $c_s^2=\Gamma P_0/\rho_0$ is the square of the adiabatic sound speed.

We imposed boundary conditions including the regularity of the gravitational perturbations, zero radial velocity on the rigid inner boundary, and vanishing Lagrange pressure perturbation at the surface, that is $\delta P=P'+u_r'/(-i \omega)\nabla P_0=0$. In terms of $h'$ and $u_r'$, the last boundary condition can be written as  \citep{Dewberry2021PSJ}
\begin{equation}
   \left ( -i \omega h'-g u_r' \right ) |_{r=R}=0.
\end{equation}
As the viscous force was included, additional boundary conditions were required to complete the boundary value problem. We used the so-called stress-free conditions, in other words the tangential stresses vanish at both boundaries.

For a given tidal potential $\Psi_l^m=\mathcal A (r/R)^l Y_l^m(\theta,\phi) \mathrm{e}^{-\mathrm{i}\omega t}$,  the tidal perturbations (including both equilibrium and dynamical tides) $\bm u'$, $h'$ and $\Phi'$ can be expanded as
\begin{equation} \label{eq:u_nm}
\bm u'= \sum  _{n=m}^L u_n^m(r) \bm R_n^m+\sum  _{n=m}^L v_n^m(r) \bm S_n^m+ \sum   _{n=m}^L w_n^m(r) \bm T_n^m ,
\end{equation}
\begin{equation}\label{eq:h_nm}
    h'=\sum _{n=m}^L h_n^m(r)Y_n^m(\theta,\phi),
\end{equation}
\begin{equation} \label{eq:phi_nm}
    \Phi'=\sum  _{n=m}^L \Phi_n^m(r)Y_n^m(\theta,\phi),
\end{equation}
where $\bm R_n^m$, $\bm S_n^m$, $\bm T_n^m$ are vector spherical harmonics
\begin{equation}
\bm R_n^m=Y_n^m(\theta,\phi) \bm{\hat{r}}, \quad \bm S_n^m=r \bm{\nabla} Y_n^m(\theta,\phi) , \quad  \bm T_n^m=r \bm{\nabla \times}  \bm R_n^m.
\end{equation}
As the basic state is axisymmetric, the perturbations involve spherical harmonics with the same order $m$ as the tidal potential $\Psi_l^m$, but the Coriolis force would couple all spherical harmonics with degree $n \ge m$. For numerical calculations, we had to make a truncation at certain degree $L$.  Substituting expansions of Eqs. (\ref{eq:u_nm}-\ref{eq:phi_nm}) into Eqs. (\ref{eq:u}-\ref{eq:phi}) and projecting onto spherical harmonics, we ended up with a set of ordinary differential equations (ODEs) involving $u_n^m(r)$, $v_n^m(r)$, $w_n^m(r)$, $h_n^m(r)$, and $\Phi_n^m(r)$. For the radial dependence, we used Chebyshev collocation on $N_r$ Gauss–Lobatto nodes \citep{Rieutord2001JFM}. The boundary conditions were applied through replacing the ODEs with the corresponding boundary conditions on the boundary nodes. The regularity of gravitational perturbations requires
\begin{equation} 
r \frac{d \Phi_n^m}{dr}+(n+1) \Phi_n^m=0 \quad \mathrm{at} \ r=R,
\end{equation}
\begin{equation} 
r \frac{d \Phi_n^m}{dr}-n\Phi_n^m=0 \quad \mathrm{at} \ r=R_i.
\end{equation}
The vanishing Lagrangian pressure perturbation at the surface and zero radial velocity at the rigid inner boundary give
\begin{equation}
-i \omega h_n^m =g u_n^m \quad \mathrm{at} \ r=R,
\end{equation}
\begin{equation}
u_n^m =0 \quad \mathrm{at} \ r=R_i.
\end{equation}
The stress-free boundary condition is given as \citep{Ogilvie2009MNRAS}
\begin{equation}
u_n^m+r\frac{d v_n^m}{dr}-v_n^m=0,\quad  r\frac{d w_n^m}{dr}-w_n^m=0\end{equation}
at both boundaries.

Using the numerical discretization described above, the boundary value problem becomes a linear system involving a large complex block-tridiagonal matrix. The solution of the linear system was obtained using the standard direct solver.  We used typical truncations of $L=200$ and $N_r=100$ for this study.

Once the solution of the linear system is obtained numerically, the complex tidal Love number is readily given by
\begin{equation}
K_l^m=\Phi_l^m(r=R)\end{equation}
for the tidal potential component $\Psi_l^m=\mathcal A (r/R)^l Y_l^m(\theta,\phi) \mathrm{e}^{-\mathrm{i}\omega t}$ (we simply set $\mathcal A=1$ for the linear tidal response). We note that the solution includes both the equilibrium and dynamical tides. For the real part of Love numbers, of particular interest is the fractional correction of dynamical tides
\begin{equation}
\Delta k_{lm}=(k_{lm}-k_{lm}^{(hs)})/k_{lm}^{(hs)},
\end{equation}
where $k_{lm}^{(hs)}$ is the hydrostatic value and it was calculated by setting $\omega=0$. {As our calculations neglect the centrifugal effect which significantly influences the high-degree Love number $k_{42}$, the calculated value of $\Delta k_{42}$ should be multiplied by the factor 0.07 when compared with the observation as we have discussed in Sec. \ref{sec:eq}.}

We can also calculate the tidal dissipation rate $D_\nu$ from the velocity perturbations  
\begin{equation}
D_\nu=\int_V 2 \mu \bm S^2 d V,
\end{equation}
where the integral was taken over the fluid domain. The dissipation rate is related to the imaginary part of the tidal Love number \citep{Ogilvie2014}
\begin{equation}
D_\nu=\frac{(2l+1)R\mathcal{A}^2}{8\pi G}\omega \mathrm{Im}[K_l^m],
\end{equation}
which can be used as an independent validation of the numerical code. The above relation is satisfied to a high degree of accuracy for all of numerical calculations presented in this paper.

\subsection{Interior models}
\begin{figure*}[!h]
   \begin{center}
      \includegraphics[width=0.9\textwidth]{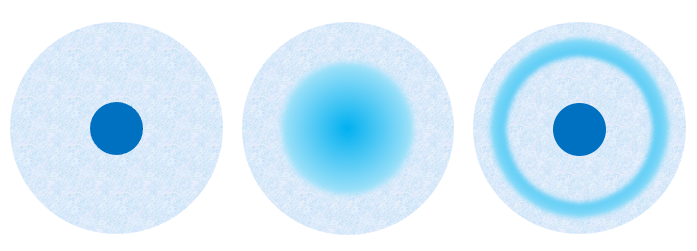}
      
    \includegraphics[width=0.9\textwidth]{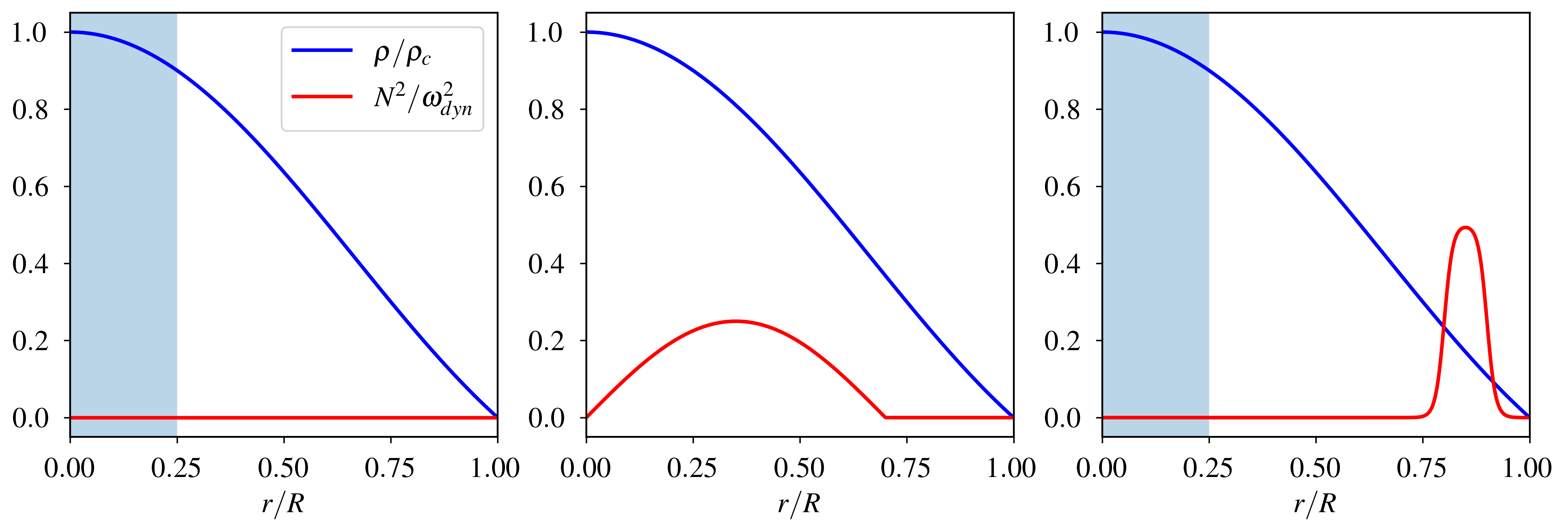} 
    
    (a) \hspace*{4cm} (b) \hspace*{4cm} (c)
      \caption{Three nominal models of Jupiter's interior used in this study. The top panel shows the schematic models and the bottom panel shows the density (normalized by the density at the center) and the Brunt-V\"ais\"al\"a frequency (normalized by the dynamical frequency) as a function of the radius. The blue shadow in the bottom panel indicates solid regions. (a) A compact rigid core model; (b) an extended dilute core model; and (c) a compact rigid core and an outer stable layer model.}
         \label{fig:model}
         \end{center}
   \end{figure*}
%----------------------------------------------------------------- 
 In order to solve Eqs. (\ref{eq:u}-\ref{eq:phi}), we need to prescribe basic state profiles $\rho_0(r)$, $g(r)$, and $N^2(r)$ to model Jupiter's interior.
Our understanding of Jupiter's interior has been significantly improved by Juno observations \citep{Stevenson2020}, yet some degrees of uncertainty remain. In this study, we do not aim to build a realistic model of Jupiter's interior, but focus on the fractional contributions of dynamical tides to the tidal Love number for different possible scenarios of Jupiter's interior. We consider three nominal interior models (Fig. \ref{fig:model}) based on a polytrope of index 1, which is a good leading order approximation for Jupiter \citep{Stevenson2020}. 

For all of models used in this study, the unperturbed density and gravity follow a hydrostatic polytrope of index 1, 
\begin{equation}
\rho_0=\frac{\pi M}{4 R^3}\frac{\sin kr}{kr},
\end{equation}
 \begin{equation}
g=\frac{G M}{ r^2}\left[\sin (kr) -kr\cos(kr)\right],
\end{equation}
where $k=\pi/R$.
The first model consists of a small rigid core of radius $0.25R$ and an isentropic fluid envelope, that is $\Gamma=2$ and $N^2=0$ in the fluid region (Fig. \ref{fig:model}(a)). 

The second model assumes an extended dilute core of radius $0.7R$ and an isentropic envelope (Fig. \ref{fig:model}(b)). The dilute core is treated as a stably stratified fluid layer with the Brunt-V\"ais\"al\"a frequency given by
\begin{equation}
\frac{N^2}{\omega_{dyn}^2}=\tilde{N}_2\sin\left(\frac{\pi r}{R_c} \right),
\end{equation}
where $\tilde{N}_2=0.25$ and $R_c=0.7$ for this model. As we fixed the density and pressure profiles to that of a polytrope, the stratification was effectively realized by adjusting the adiabatic index $(\Gamma>2)$ in the dilute core \citep{Lai2021PSJ}. This model is similar to the one used in \cite{Idini2022PSJ}, but they adjusted the density profile to model the stable stratification in the dilute core while fixing the adiabatic index $\Gamma=2$.

The third model is based on the model in Fig. \ref{fig:model}(a), but we further added a stably stratified layer between $0.8R$ and $0.9R$ (Fig. \ref{fig:model}(c)), possibly resulting from  H-He immiscibility \citep{Debras2019ApJ,Stevenson2022PSJ}. The Brunt-V\"ais\"al\"a frequency in the top stable layer is prescribed as
\begin{equation}
\frac{N^2}{\omega_{dyn}^2}=\tilde{N}_2\frac{1}{[1+ \mathrm e^{-100(r-0.8)}][1+ \mathrm e^{100(r-0.9)}]}.
\end{equation}
The degree of stratification of this layer remains uncertain, but it is estimated that typical values of $N^2/{\omega_{dyn}^2}$ would be roughly between 0.1 and 0.8 for Jupiter \citep{Christensen2020ApJ,Gastine2021Icar}. Here we set a moderate value $\tilde{N}_2=0.5$. 
 
 We note that an interior model with the coexistence of a dilute core and a top stable layer is also possible \citep{Debras2019ApJ}. As this kind of model involves two different stably stratified layers, it would be difficult to characterize the role of the top stable layer on tides. We considered only the combination of a compact rigid core and a top stable layer for simplicity.
 
In all of these models, we set the total mass $M$, the radius $R$, and the spin rate $\Omega$ such that the ratio $\epsilon=\Omega/\sqrt{GM/R^3}=0.288$, corresponding the value of Jupiter. Our calculations also require the fluid viscosity, which is difficult to estimate in detail for giant planets. We simply assumed the dynamic viscosity $\mu$ is proportional to the background density $\rho_0$, so the kinematic viscosity $\nu=\mu/\rho_0$ is constant. The viscosity can be measured by the dimensionless number $Ek=\nu/(\Omega R^2)$, known as the  Ekman number. We set $Ek=10^{-6}$ for most of the calculations (unless otherwise specified), roughly corresponding to the effective viscosity based on mixing-length theory  \citep{Guillot2004}.  

As we have mentioned that we do not aim to construct a realistic interior model for Jupiter in this study.  These simplified models were designed to investigate the effects of a compact rigid core, an extended dilute core, and a top stable layer on the tidal responses of Jupiter. Nevertheless, the fractional corrections $\Delta k_{lm}$ and the tidal quality factor $Q$ for these simplified models can be used to make some qualitative comparisons with the observations \citep{Lai2021PSJ,Idini2021PSJ,Idini2022PSJ}. 

\section{Results} \label{sec:Res}
In this paper, we focus on the dominant tidal component $\Psi_2^2$ and a high-degree tesseral component $\Psi_4^2$, for which non-negligible dynamical corrections have been detected as we have discussed in Sec. \ref{sec:Intro}.  Our calculations are limited to the frequency range of $-2\le \omega/\Omega\le -1$, which is relevant to the tidal frequencies of the Galilean moons. The negative tidal frequency means that the tidal forcing is retrograde in the corotating frame with the planet based on our convention. For the real part of Love numbers, we show the fractional correction $\Delta k_{lm}$. {In order to make comparisons with the Juno observation, the calculated  $\Delta k_{42}$ was multiplied by 0.07 to compensate for the centrifugal effect which is neglected in our calculations. }  
Because of the negative tidal frequency, the imaginary part of Love numbers is also negative in our calculations and is related to the tidal quality factor by $k_{lm}/Q_l=-\mathrm{Im}[K_l^m]$ according to Eq. (\ref{eq:Q}). 

\subsection{Full polytrope model}
\begin{figure*}[!h]
\centering
\includegraphics[width=0.8\textwidth]{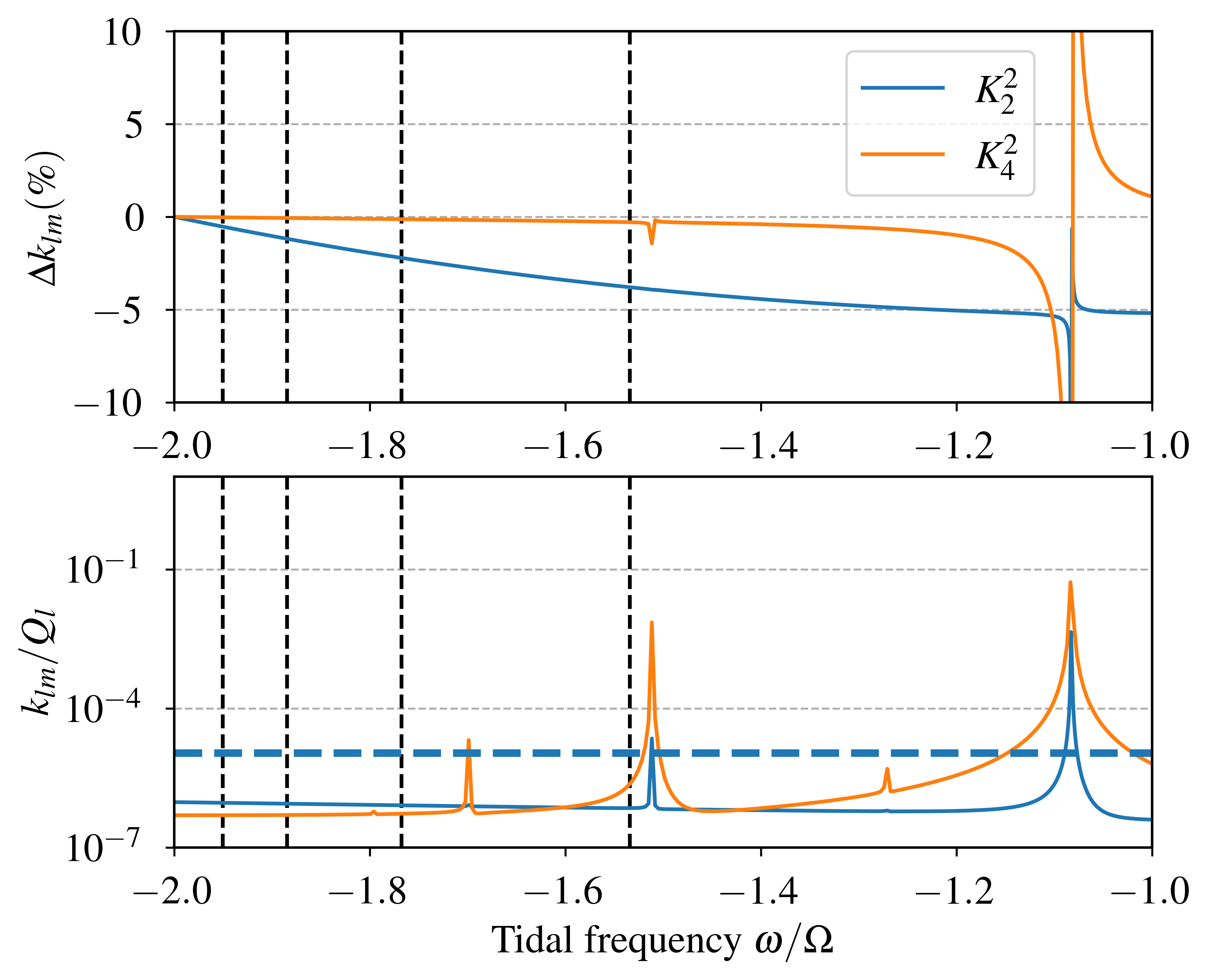}
\caption{Complex Love number as a function of the tidal frequency for a full isentropic polytrope of index 1. The top panel shows the fractional correction $\Delta k_{lm}$ of the real part of  the Love numbers. {The fractional correction $\Delta k_{42}$ (orange curve in the top panel) was multiplied by 0.07. }The bottom panel shows the minus imaginary part $-\mathrm{Im}[K_l^m]$, which is equivalent to $k_{lm}/Q_l$. Vertical dashed lines indicate tidal frequencies of four Galilean moons of Jupiter (from right to left: Io, Europa, Ganymede, and Callisto). The horizontal dashed line in the bottom panel represents the astrometric observation of the frequency independent $k_2/Q_2$ from \cite{Lainey2009}.}
 \label{fig:LoveKModel0}
\end{figure*}
\begin{figure*}[!h]
\centering
\includegraphics[width=0.4\textwidth]{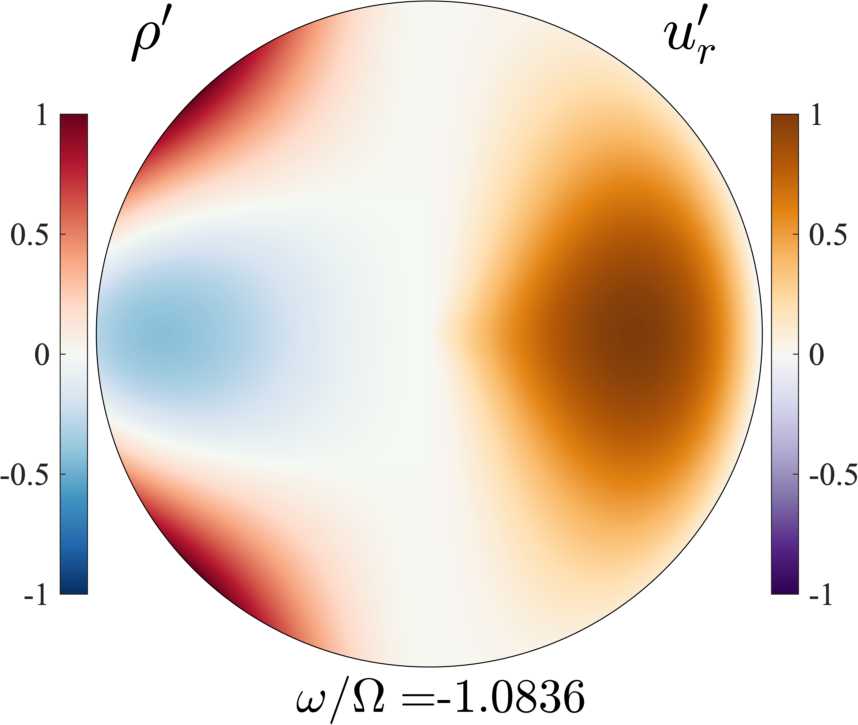}
\hspace{1cm}
\includegraphics[width=0.4\textwidth]{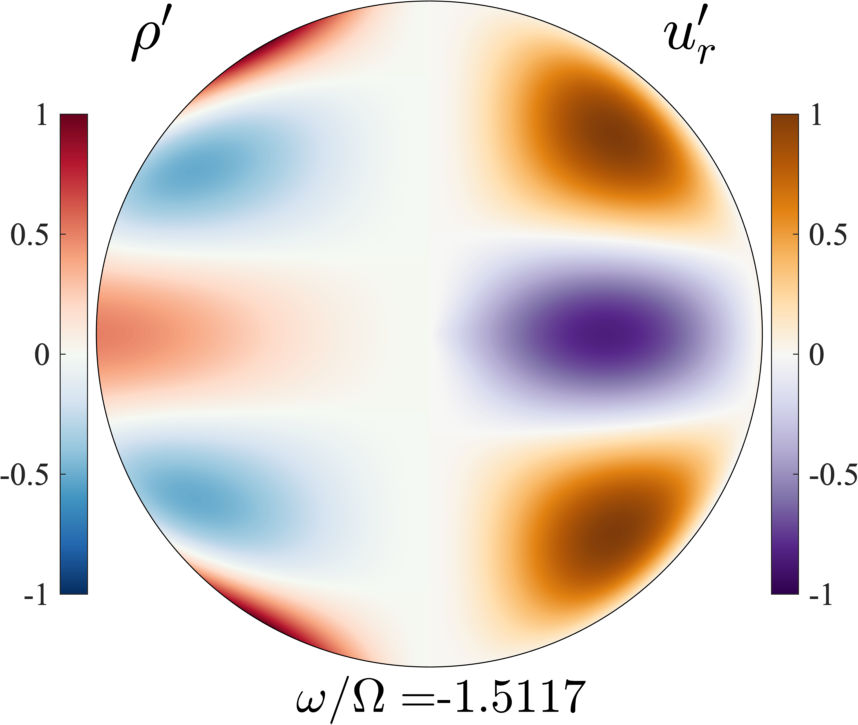} \\
(a) \hspace{7cm} (b)
\caption{Density perturbations (left half) and radial velocity perturbations (right half) in the meridional plane to the tidal component $\Psi_4^2$ at two resonant  frequencies in Fig. \ref{fig:LoveKModel0}.  Amplitudes were normalized by the maximum absolute values.}
 \label{fig:Mode0}
\end{figure*}

Before presenting results for the interior models in Fig. \ref{fig:model}, we first show the tidal response  of a full isentropic polytrope, that is to say neutrally buoyant in the whole fluid sphere. This model serves as a reference for other models and has been used to investigate the dynamical tides of Jupiter in recent analytical studies \citep{Idini2021PSJ,Lai2021PSJ}.  Fig. \ref{fig:LoveKModel0} shows both the real and imaginary parts of the Love numbers as a function of the tidal frequency for the full polytrope model. We can see that  $\Delta k_{22}$ is negative in the frequency range we considered and it smoothly varies as the tidal frequency except at a burst around $\omega/\Omega=-1.08$, which corresponds to a resonance with an inertial mode. Away from resonances, our numerical results are consistent with recent theoretical calculations and produce $\Delta k_{22}\approx-4\%$  at the tidal frequency of Io \citep{Lai2021PSJ,Idini2021PSJ}.  These studies  also revealed that the dynamical correction $\Delta k_{22}$ can be attributed to the Coriolis effect on the $f$-modes.  Apart from the $f$-modes, the rotating sphere of isentropic fluid also supports smooth inertial modes restored by the Coriolis force in the frequency range of $0<|\omega/\Omega|<2$ \citep{Greenspan1968,Lockitch1999ApJ}. The burst of $\Delta k_{22}$ at $\omega/\Omega=-1.08$  is indeed due to the resonant excitation of the inertial mode as shown in Fig. \ref{fig:Mode0}(a), but we noticed that the resonance occurs only in a very narrow frequency range.

{However, this inertial mode  has more significant contributions to $\Delta k_{42}$.} The angular structure of an inertial mode  cannot be described by single spherical harmonics in general \citep{Lockitch1999ApJ}, but the density perturbations (and thus the gravitational perturbations) are dominated by the spherical harmonics $Y_4^2(\theta,\phi)$ for the resonant inertial mode at $\omega/\Omega=-1.0836$ as we can see from Fig. \ref{fig:Mode0}(a). This suggests a likely strong coupling between the tidal potential component $\Psi_4^2$ and the inertial mode in Fig. \ref{fig:Mode0}(a), that is to say large tidal overlap as described in \cite{Wu2005ApJII}, leading to significant dynamical corrections to $k_{42}$.
{The dynamical correction can reach $\Delta k_{42}\approx -10\%$ (after the centrifugal correction) near the resonance at $\omega/\Omega=-1.0836$. However, the tidal frequencies of the Galilean satellites are too far away from this resonance.}

The curve of $\Delta k_{42}$ also shows a spike around $\omega/\Omega=-1.51$, corresponding to a narrow resonance with a high-degree inertial mode ($\rho'$ is dominated by $Y_6^2(\theta,\phi)$ as shown in Fig. \ref{fig:Mode0}(b)). Interestingly, the tidal frequency of Io is close to this resonance, but the dynamical correction caused by this resonant mode is insufficient to account for the observed $\Delta k_{42}\approx -11\%$. The frequencies of inertial modes in Fig. \ref{fig:Mode0} are slightly shifted compared to the results of \cite{Lockitch1999ApJ} for a polytrope of index 1 (see their table 6 and note the different conventions for the sign of frequencies) because they  assumed $\epsilon \to 0$ whereas we set $\epsilon=0.288$.

The imaginary parts of the Love numbers in Fig. \ref{fig:LoveKModel0}  show that resonances with inertial modes significantly enhance the tidal dissipation. The enhanced dissipation due to resonant inertial modes  in a neutrally buoyant sphere has been demonstrated by \cite{Wu2005ApJII}, but using different density profiles. When the tidal frequency is away from resonances, the dissipation rate for the full isentropic polytrope is too small to account for the observed tidal quality factor $Q$ \citep{Lainey2009}.

\subsection{Compact rigid core model}
\begin{figure*}[!h]
\centering
\includegraphics[width=0.8\textwidth]{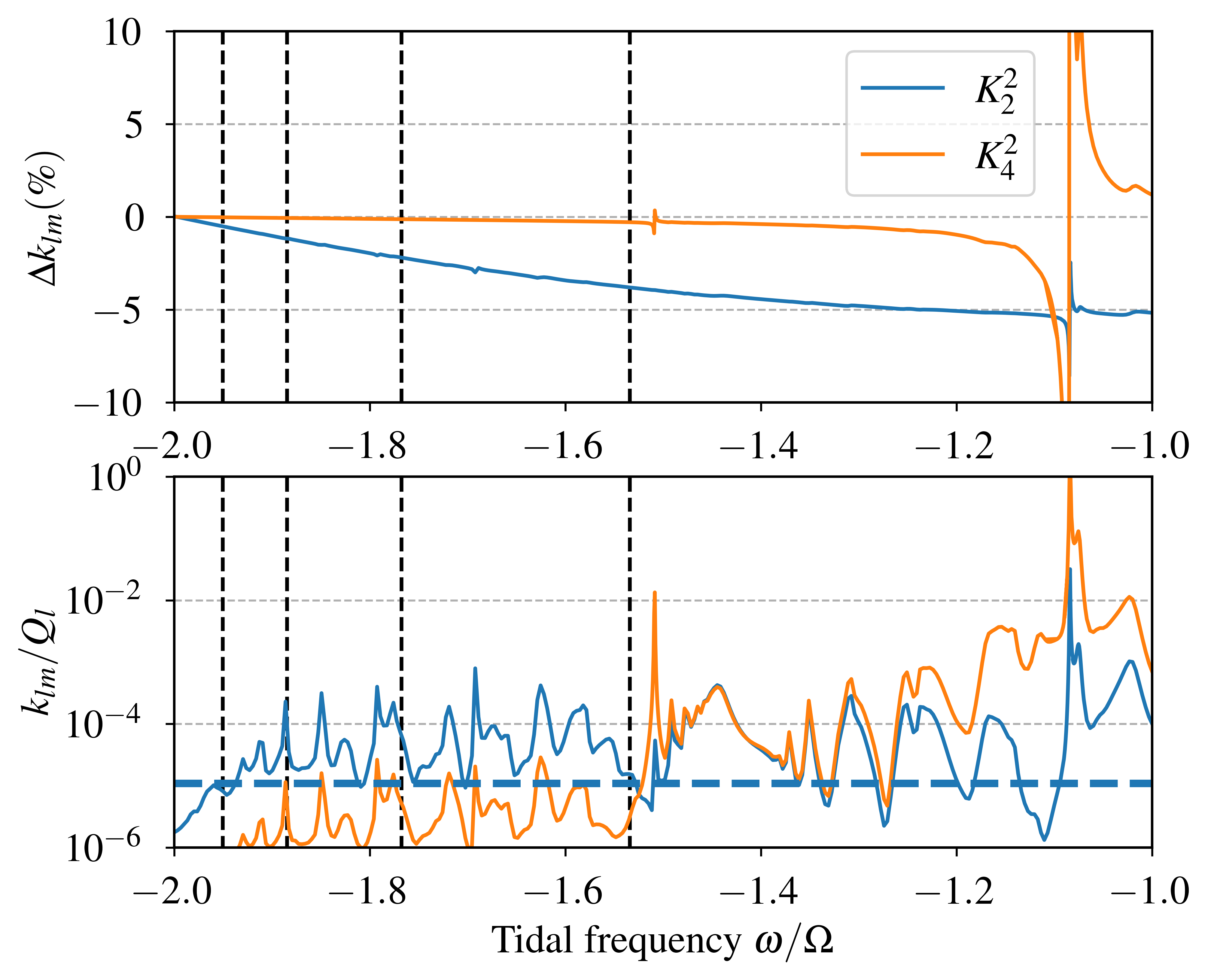}
\caption{As for Fig. \ref{fig:LoveKModel0}, but for the interior model with a compact rigid core. {The fractional correction $\Delta k_{42}$ (orange curve in the top panel) was multiplied by 0.07. } }
 \label{fig:LoveKModel1}
\end{figure*}

\begin{figure*}[!h]
\centering
\includegraphics[width=0.4\textwidth]{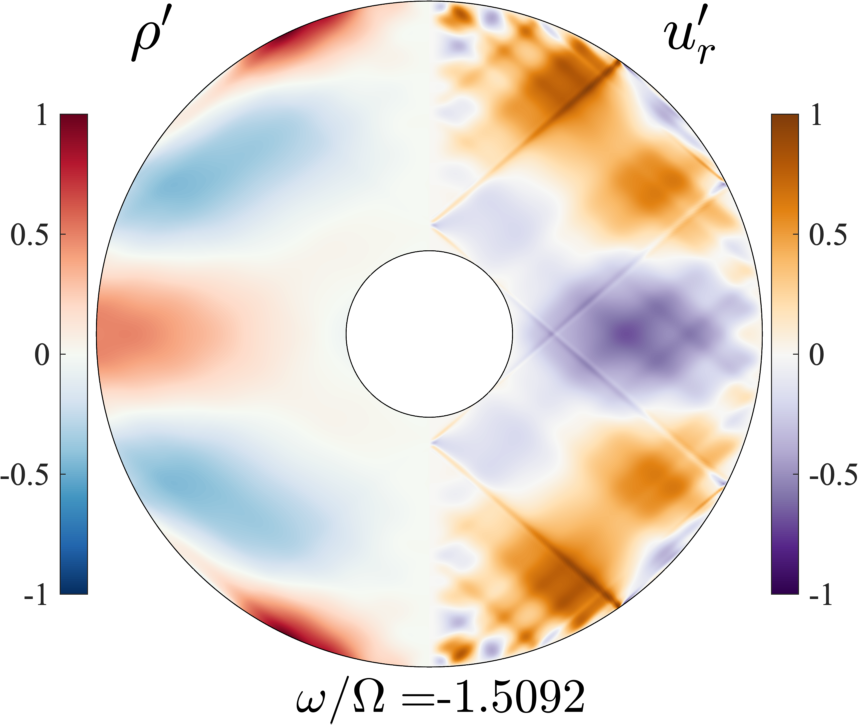}
\hspace{1cm}
\includegraphics[width=0.4\textwidth]{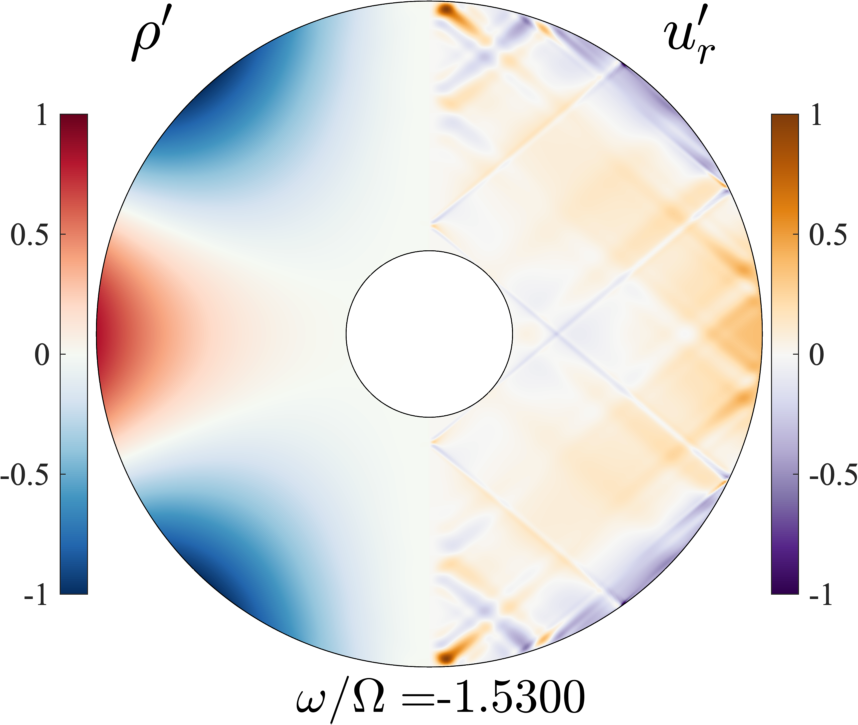} \\
(a) \hspace{7cm} (b)
\caption{Density perturbations (left half) and gravitational perturbations (right half) in the meridional plane to the tidal component $\Psi_4^2$ for the interior model with a compact rigid core at (a) $\omega/\Omega=-1.5092$ (resonance) and (b) $\omega/\Omega=1.53$ (nonresonance) with $Ek=10^{-7}$.  Amplitudes were normalized by the maximum absolute values.}
 \label{fig:Mode1}
\end{figure*}

We now consider tidal responses for the interior model with a compact rigid core. Basically, the inner region ($r\le 0.25R$) of a whole fluid polytrope becomes solid for this model. 
Fig. \ref{fig:LoveKModel1} shows the frequency dependence of the Love numbers for the compact rigid core model. We can see that the real parts are largely similar to those of a full polytrope, but the imaginary parts are rather different from those of a full polytrope, showing enhanced tidal dissipation by introducing the rigid core. The rigid core model also supports inertial waves in the fluid envelope, but these waves have some peculiar behaviors due to the singularity in a spherical shell \citep{Stewartson1969}.  Smooth inertial modes generally do not  exist in a spherical shell even with uniform density \citep{Rieutord2001JFM}, and localized wave beams spawned from the critical latitudes propagate in the bulk along the characteristics of the inertial wave equations \citep[e.g.,][]{Ogilvie2009MNRAS}. However, \cite{Lin2021ApJ} recently revealed that resonant tidal responses in a spherical shell correspond to eigen modes with large-scale flows hidden beneath localized wave beams using a uniform density model. Furthermore, it was shown that the hidden large-scale structures basically resemble inertial modes in a full sphere.  This is in line with our results for the nonuniform density model in this study. The real parts, $k_{22}$ and $k_{42}$,  are relevant to only large-scale density perturbations, which are similar to  inertial modes in a full sphere as one can see from Fig. \ref{fig:Mode1}. Therefore, the curves of $\Delta k_{lm}$ for the rigid core model resemble those of a full polytrope, but we note slight shifts of the resonant frequencies due to the presence of a rigid core. {As for the full polytrope, the compact rigid core model can produce $\Delta k_{22}=-4\%$ as observed, but it cannot produce sufficient dynamical correction in the high-degree Love number $k_{42}$ near the tidal frequency of Io to account for the observed $\Delta k_{42}=-11\%$. }

On the other hand, the imaginary parts are largely modified by the presence of a small rigid core. We can see that the tidal dissipation is significantly enhanced by the localized wave beams spawned from the critical latitudes both  in and out of resonances.  
The velocity perturbations in  Fig. \ref{fig:Mode1}(b) indeed exhibit localized waves propagating in the bulk, which can generate significant viscous dissipation but they do not produce significant density and gravitational perturbations. In Fig. \ref{fig:LoveKModel1}, we also see that several peaks in the tidal dissipation (bottom panel) do not lead to obvious fluctuations in $\Delta k_{lm}$ (top panel), corresponding to resonances with higher degree modes that contribute little to the low degree (i.e., $l=2$ and $l=4$) gravitational perturbations. 

In summary for the compact rigid core model, the tidal dissipation is significantly enhanced with respect to the full polytrope case. This is in line with the early work of \cite{Ogilvie2004ApJ}, who showed the enhanced tidal dissipation due to inertial waves in the convective envelope of rotating stars and planets. The averaged dissipation in the tidal frequency range of Galilean moons  gives rise to a comparable tidal quality factor as observed by \cite{Lainey2009}. However, the fractional correction to the real part of Love number $\Delta k_{42}$  is insufficient to explain the observation.

\subsection{Dilute core model}
\begin{figure*}[!h]
\centering
\includegraphics[width=0.8\textwidth]{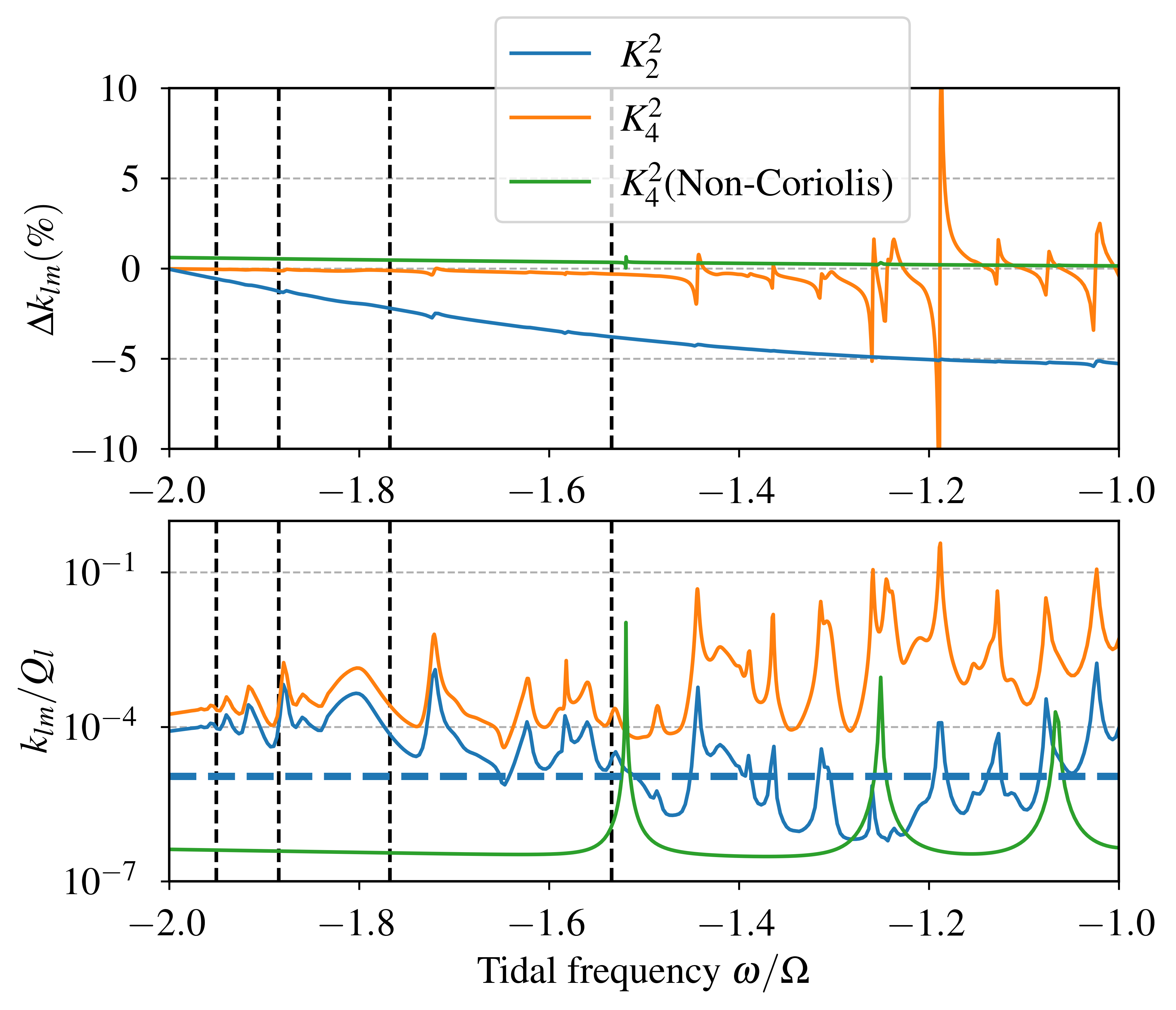}
\caption{As for Fig. \ref{fig:LoveKModel0}, but for the interior model with an extended dilute core. Green lines represent results without including the Coriolis force. {The fractional correction $\Delta k_{42}$ (orange and green curves in the top panel) was multiplied by 0.07. }}
 \label{fig:LoveKModel2}
\end{figure*}

\begin{figure*}[!h]
\centering
\includegraphics[width=0.4\textwidth]{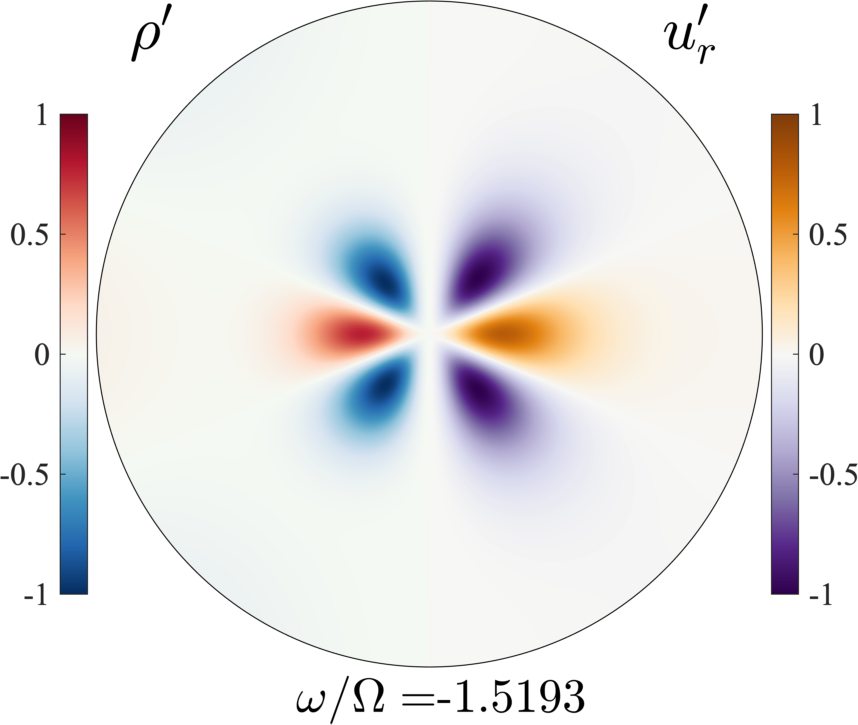}
\hspace{1cm}
\includegraphics[width=0.4\textwidth]{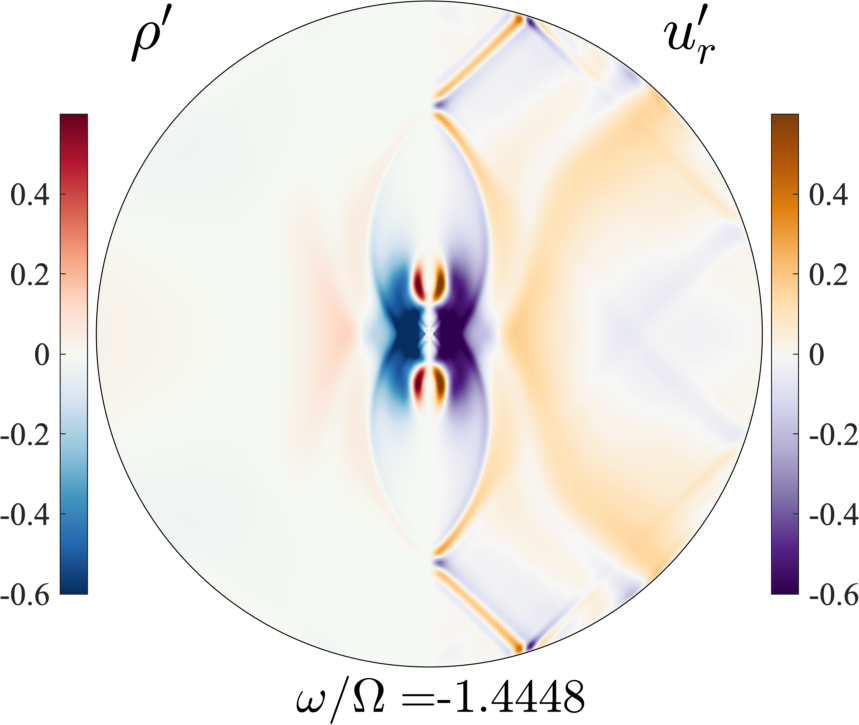} \\
(a) \hspace{7cm} (b)
\caption{Density perturbations (left half) and radial velocity perturbations (right half) in the meridional plane to the tidal component $\Psi_4^2$ for the interior model with an extended dilute core. (a) Without including the Coriolis force at $\omega/\Omega=-1.5193$ (resonance); (b) including the Coriolis force at $\omega/\Omega=-1.4448$ (resonance).  Amplitudes were normalized by the maximum absolute values.}
 \label{fig:Mode2}
\end{figure*}

An extended dilute core rather than a compact core in Jupiter has been suggested recently based on Juno gravitational measurements \citep{Wahl2017GeoRL,Militzer2022PSJ}. In this subsection, we consider tidal responses for the interior model with an extended dilute core as shown in Fig. \ref{fig:model}(b). The dilute core was treated as a stably stratified layer which supports gravity waves restored by the buoyancy. If the Coriolis force is fully taken into account, dynamical tides in the dilute core region would be in the form of gravito-inertial waves \citep{Dintrans1999,Xu2017PhRvD}.  \cite{Idini2022PSJ} recently calculated the tidal response of Jupiter with an extended dilute core, but they did not fully consider the Coriolis effect, which turns out to be important as we subsequently show.      

Fig. \ref{fig:LoveKModel2} shows the frequency dependence of the Love numbers for the dilute core model.  For the tidal component $\Psi_2^2$ (blue curves), the dynamical correction $\Delta k_{22}$ is generally similar to that of the full polytrope, except for the absence of obvious spikes in the dilute core model. However, the imaginary part exhibits several peaks and troughs, suggesting possible resonances with high-degree mixed modes that enhance the tidal dissipation but do not significantly contribute to the $l=2$ gravitational perturbations. The overall tidal dissipation is also enhanced with respect to the full polytrope due to the excitation of gravito-inertial waves in the dilute core and inertial waves in the convective envelope. The frequency-averaged tidal dissipation tends to be compatible with the observed tidal quality factor as we can see from Fig. \ref{fig:LoveKModel2}.

For the tidal component $\Psi_4^2$,  Fig. \ref{fig:LoveKModel2} also shows results without including the Coriolis force (green curves) for comparison. We  note that the fractional correction $\Delta k_{42}$ is always positive when the Coriolis force is neglected, probably because the pure gravity modes enhance the in-phase gravitational perturbations and thus produce positive dynamical corrections. Nevertheless, we observed distinct resonant responses at certain tidal frequencies from both real and imaginary parts of the Love number for the non-Coriolis case. For instance, the resonance at $\omega/\Omega=-1.5193$, which is close to the tidal frequency of Io, corresponds to the first gravity mode of $l=4$ and $m=2$ as shown in Fig. \ref{fig:Mode2} (a).  Indeed, \cite{Idini2022PSJ} propose the resonant locking between this gravity mode  \footnote{They used slightly different background density $\rho_0(r)$ and  Brunt-V\"ais\"al\"a frequency $N(r)$, so the mode frequency is slightly shifted.} (referred to as $_4^2g_1$) and the Jupiter-Io orbital evolution to explain the observed $\Delta k_{42}$ for Jupiter. In \cite{Idini2022PSJ},  the Coriolis force was neglected for the calculation of gravity modes, but approximated rotational  corrections were made to obtain the Love number. 
However, fully taking the Coriolis force into account significantly alters the tidal responses as we can see from Fig. \ref{fig:LoveKModel2}. {The dynamical correction $\Delta k_{42}$ exhibits several large fluctuations especially in the frequency range of $-1.5<\omega/\Omega <-1$. This is due to the mixing of gravity modes and inertial modes in the dilute core, leading to more chances for resonances. The most significant dynamical corrections are produced near the tidal frequency $\omega/\Omega=-1.2$, which is close to the frequency of the purely inertial mode as shown in Fig. \ref{fig:Mode0}(a). Of course,  the inertial mode is mixed with gravity modes in the dilute core for this model. The resonance around $\omega/\Omega=-1.2$ can produce more than $-10\%$ dynamical corrections in $k_{42}$ (after the centrifugal correction), but it is too far away from the tidal frequency of Io. The resonance close to the tidal frequency of Io (also close to the frequency of pure gravity mode  $_4^2g_1$) occurs at $\omega/\Omega=-1.4448$ when the Coriolis force is considered.  } Fig. \ref{fig:Mode2} (b) shows the spatial structure of this resonant response. The Coriolis effect not only leads to a non-negligible shift in the mode frequency, but also largely modifies the mode structure. The perturbations are in the from of gravito-inertial waves in the dilute core and become pure inertial waves in the neutrally buoyant envelope. Non-negligible dynamical corrections are induced by this resonance at $\omega/\Omega=-1.4448$, but the corrections are insufficient (after the centrifugal correction) to account for the observed $\Delta k_{42}=-11\%$. {As the resonance is very narrow, we used 200 equally spaced frequency points in the tidal frequency interval of [-1.45, -1.43]. The peak amplitude of $\Delta k_{42}$ in this frequency interval is comparable to the amplitude in the calculations using only 20 frequency points}, suggesting that the frequency sampling points are sufficient to capture the resonant peak.

Comparing the orange and green curves in the bottom panel of Fig. \ref{fig:LoveKModel2}, we can see that the tidal dissipation is increased by about two orders of magnitude when the Coriolis force is included. This suggests that the excitation of pure gravity waves is a less efficient tidal dissipation mechanism (unless resonances take place) based on our linear calculations, though the nonlinear interaction or wave breaking of gravity waves may lead to efficient tidal dissipation \citep[e.g.,][]{Barker2011MNRAS,Weinberg2012ApJ}.

\subsection{Outer stable layer model}
\begin{figure*} [!h]
\centering
\includegraphics[width=0.8\textwidth]{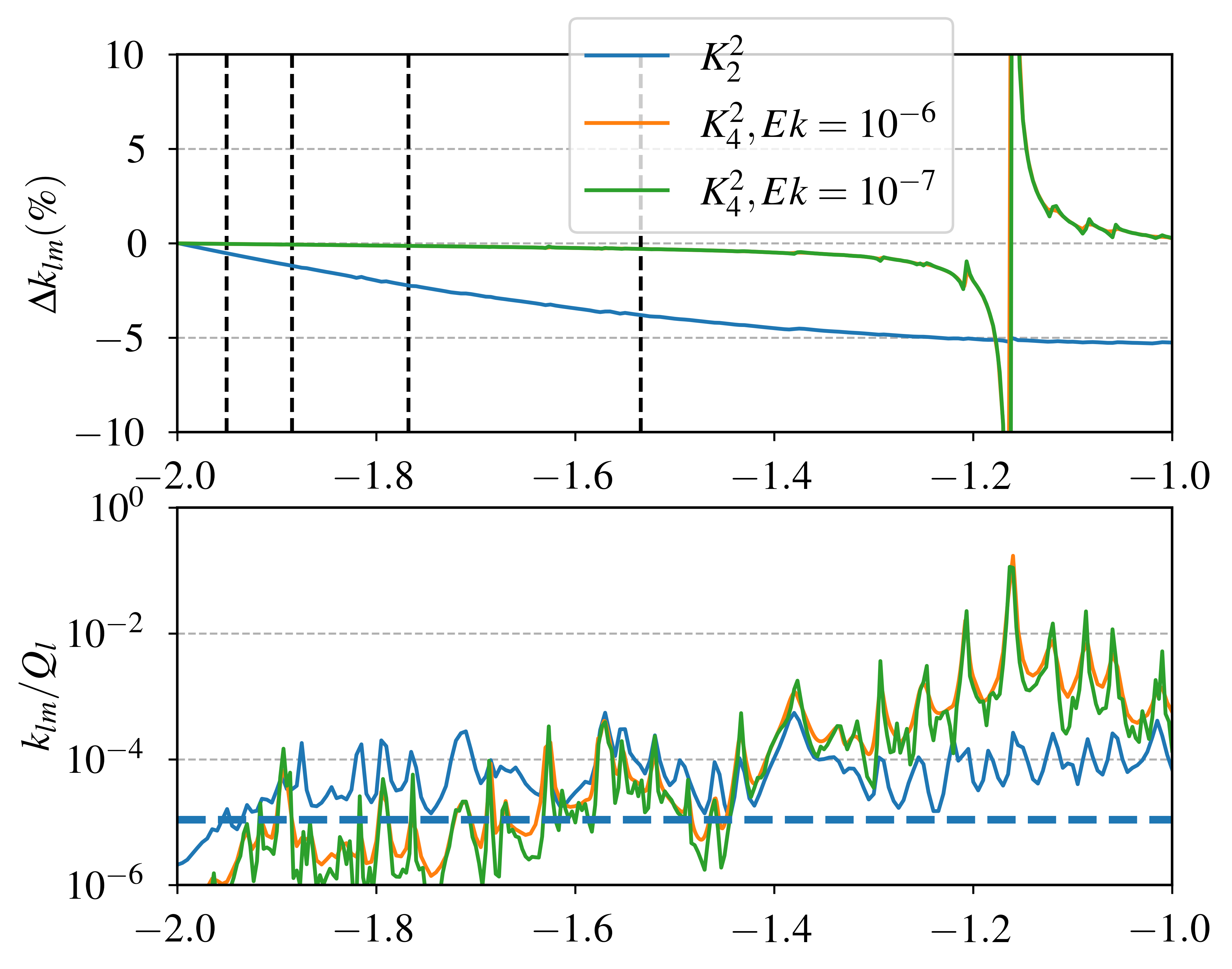}
\caption{As for Fig. \ref{fig:LoveKModel0}, but for the interior model with a small rigid core and a top stably stratified layer. Green lines represent results at the Ekman number $Ek=10^{-7}$. {The fractional correction $\Delta k_{42}$ (orange and green curves in the top panel) was multiplied by 0.07.}}
 \label{fig:LoveKModel3}
\end{figure*}

\begin{figure*} [!h]
\centering
\includegraphics[width=0.4\textwidth]{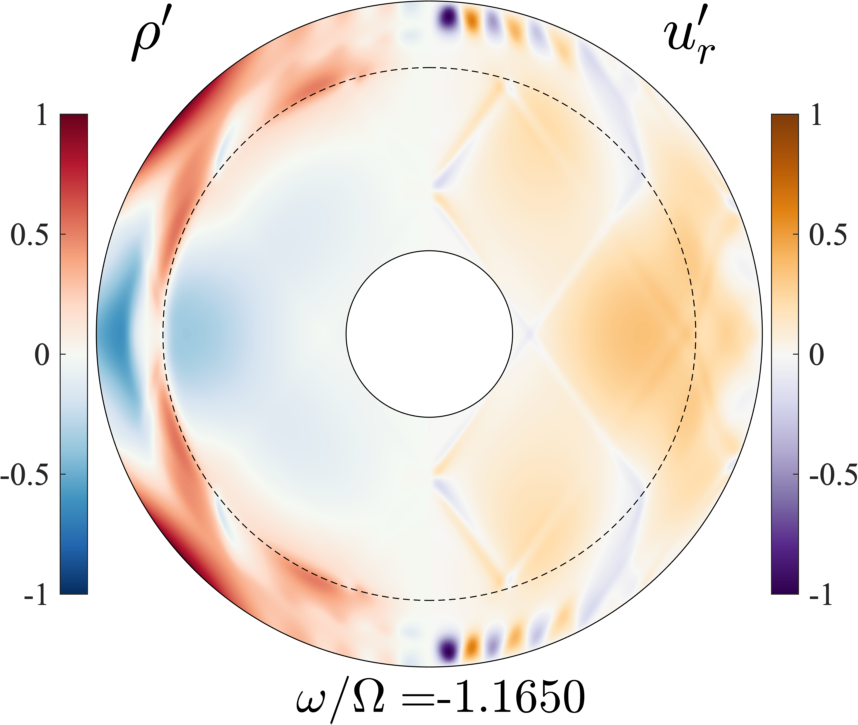}
\hspace{1cm}
\includegraphics[width=0.4\textwidth]{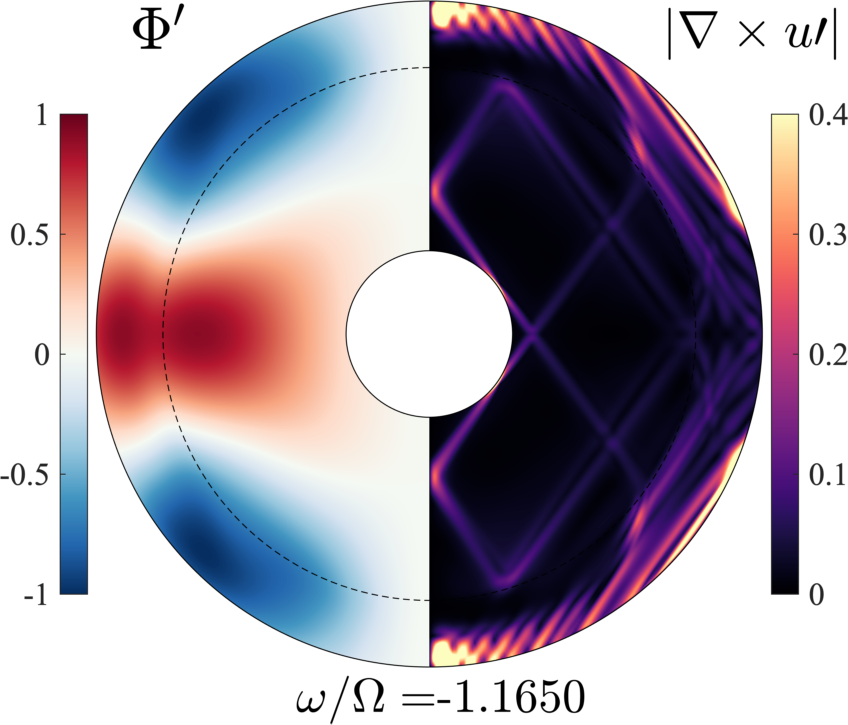} \\
(a) \hspace{7cm} (b)
\caption{Perturbations in the meridional plane to the tidal component $\Psi_4^2$ at $\omega/\Omega=-1.1650$ for the interior model (c) in Fig. \ref{fig:model}. (a) Density (left half) and radial velocity (right half) perturbations; (b) gravitational (left half) and vorticity (right half) perturbations.  Amplitudes were normalized by the maximum absolute values. The dashed lines denote $r=0.8R$.}
 \label{fig:Mode3}
\end{figure*}

We finally consider the effect of an outer stable layer, which may exist in Jupiter resulting from  H-He immiscibility \citep{Debras2019ApJ}. Fig. \ref{fig:LoveKModel3} shows the Love numbers as a function of the tidal frequency for the interior model (c) in Fig. \ref{fig:model}, which includes a compact rigid core and a top stable layer between $0.8R$ and $0.9R$.  For the tidal responses to $\Psi_2^2$, the dynamical correction $\Delta k_{22}$ is similar to the case without the stable layer, but the presence of the thin stable layer eliminates the spike due to the resonant inertial mode at the tidal frequency around $\omega/\Omega=-1.08$.
The overall tidal dissipation due to $\Psi_2^2$ is comparable to the counterpart without the top stable layer (blue curve in the bottom panel of Fig. \ref{fig:LoveKModel1}), but the fluctuation amplitudes (the differences between peaks and troughs) are smaller.  

For the tidal responses to $\Psi_4^2$, we also show results for $Ek=10^{-7}$ (green curves) to illustrate the effect of fluid viscosity in Fig. \ref{fig:LoveKModel3}. One can see that the viscosity has little influence on the real part of the Love number. The tidal dissipation weakly depends on viscosity at peaks and troughs, but the overall dissipation tends to be insensitive to viscosity. Indeed, \cite{Ogilvie2013MNRAS} has shown that the frequency-averaged dissipation is independent of viscosity.

{The dynamical correction $\Delta k_{42}$ is also similar to the case without the stable layer. We can see large variations in $\Delta k_{42}$  at the tidal frequency around $\omega/\Omega=-1.165$, which corresponds to a resonant mode as shown in Fig. \ref{fig:Mode3}. }
This mode is complicated because it involves three different layers for the interior model considered here. The fluid body is primarily neutrally buoyant and supports inertial waves. However, the fluid domain is separated by the thin stable layer, which suppresses radial fluid motions and creates a "barrier" for the communication between inertial waves in the inner and outer regions (see the radial velocity and vorticity perturbations in Fig. \ref{fig:Mode3}). In addition, the thin stable layer supports rotationally modified gravity waves. The density perturbations are mainly restricted in the stable layer and the outer envelope ($r>0.8R$). Despite the complicated velocity and density perturbations, the gravitational perturbations are dominated by the $l=4$ component with relatively simple radial dependence. {In this regard, this complicated mode is relevant to the $l=4$ inertial mode without the stable layer, leading to large dynamical corrections around the tidal frequency at $\omega/\Omega\approx-1.1$ as in Fig. \ref{fig:LoveKModel1}. However, the dynamical correction $\Delta k_{42}$ is negligible after the centrifugal correction at the tidal frequency of Io.}
%It is possible to produce the observed $\Delta k_{42}\approx -11\%$ at the tidal frequency of Io by adjusting the density profiles and the Brunt-V\"ais\"al\"a frequency in the stable layer.
%, but it requires large amount of calculations to figure out possible profiles. 

%Fig. \ref{fig:Mode3} also shows the vorticity     of the velocity perturbations, which is relevant to the viscous dissipation. We can see that the dissipation would 

%

\section{Conclusions}
We have developed a numerical method for calculating the tidal responses of a compressible, self-gravitating, rotating, and viscous fluid body. We have fully taken the Coriolis force into account, but neglected the centrifugal distortion, which allowed us to solve the problem in the spherical geometry.  We used the pseudo-spectral method based on spherical harmonics in the angular directions and Chebyshev collocation in the radial direction. Different from recent studies on Jupiter's dynamical tides \citep{Lai2021PSJ,Idini2022PSJ,Dewberry2022ApJ}, we directly solved the tidally forced problem and explicitly added the fluid viscosity, which allowed us to simultaneously obtain the real and imaginary parts of the tidal Love numbers for a given planetary interior model. 
 
 In this study, we have considered three simplified interior models (Fig. \ref{fig:model}) of Jupiter based on a polytrope of index 1.  We have focused on the tidal components $\Psi_2^2$ and $\Psi_4^2$ in the frequency range of $-2 \le \omega/\Omega \le -1$, which is relevant to the tidal frequencies of Galilean moons. Our numerical results show that the dynamical correction $\Delta k_{22}$ is generally insensitive to the interior models.  All of the models we considered  can give rise to the observed $\Delta k_{22}\approx -4\%$ at the tidal frequency of Io, which is also in line with previous studies \citep{Idini2021PSJ,Lai2021PSJ}.  The tidal dissipation is significantly enhanced by the presence of a compact rigid core model or an extended dilute core with respect to the full polytrope, leading to comparable tidal quality factor $Q$ as observed by \cite{Lainey2009}. 
 %Note that although $k_{22}/Q_2$ and $k_{42}/Q_4$ are comparable in most of our calculations for the tidal frequencies of Galilean moons, the orbital evolution of the satellites should be primarily controlled by the tidal component $\Psi_2^2$ because the amplitude of  
 
 {For the tidal responses to the $\Psi_4^2$ component, all of models we considered are difficult to give rise to  $\Delta k_{42}\approx -11\%$ near the tidal frequency of Io. For the interior model with a compact rigid core, significant dynamical corrections are generated at the tidal frequency around $\omega/\Omega\approx-1.1$ due to the resonance with an inertial mode whose gravitational perturbations are dominated by the spherical harmonics of  $l=4$ and $m=2$. However, this resonance is too far away from the tidal frequencies of Galilean moons. For the interior model with an extended dilute core,  we demonstrate that the gravity modes in the dilute core can be significantly modified by the Coriolis force, leading to the mixed gravito-inertial modes.  Resonances with gravito-inertial modes in the dilute core can produce non-negligible dynamical corrections, but they are insufficient to explain the observed $\Delta k_{42}\approx -11\%$ near the tidal frequency of Io based on our simplified model.  
 We also briefly investigated the effect of a top stable layer on Jupiter's tides. The thin stable layer acts as a "barrier" and tends to restrict the density and velocity perturbations mainly in the outer envelope. However, our numerical results show that the top stable layer has little influence on the real part of tidal Love numbers.}
 
 As we have mentioned, we do not aim to construct a realistic interior model of Jupiter in this study. These simplified models were designed to characterize the tidal responses of some possible scenarios of Jupiter's interior.  Because the dynamical tides highly depend on the tidal frequency, the satellite-dependent tidal Love numbers would provide more constraints on the interior of  Jupiter \citep{Idini2022PSJ}. In addition, seismology is the most effective approach to determine the interior structure of planets, though the detection of Jupiter's oscillations remains a big challenge \citep{Gaulme2011AA}. Nevertheless, the numerical scheme we developed in this study can also be used for  theoretical calculations of oscillation modes of giant planets.  
 
 There are some caveats, which should be considered in future. First, we did not consider the centrifugal deformation in order to solve the problem in the spherical geometry. {The centrifugal effect plays a significant role in the tidal Love numbers of Jupiter, especially for the high-degree tidal components. Although we have made the centrifugal corrections when the numerical results were qualitatively compared with the observations,  both the Coriolis and centrifugal effects should be self-consistently taken into account for quantitative comparisons with the high precision observations in the future.}  Second, giant planets exhibit  differential rotations, which also influence the oscillation modes and thus tidal responses  \citep{Dewberry2021PSJ}. Finally, Jupiter has the strongest magnetic field among planets in the Solar System and mainly consists of electrically conducting fluid (metallic hydrogen), so the magnetic effects \citep{Lin2018MNRAS, Wei2022AA} should also play a part in the tides of Jupiter.

\begin{acknowledgements}
 {The author would like to thank an anonymous referee for constructive comments and Dali Kong for fruitful discussions.} This study was supported by the B-type Strategic Priority Program of the CAS (XDB41000000), National Natural Science Foundation of China (grant no. 42174215) and the preresearch project on Civil Aerospace Technologies of CNSA (D020308). Numerical calculations were performed on the Taiyi cluster supported by the Center for Computational Science and Engineering of Southern University of Science and Technology.
 \end{acknowledgements}

 %\bibliographystyle{aa} % style aa.bst
 %\bibliography{JupiterTides} % your references Yourfile.bib

\begin{thebibliography}{38}
\expandafter\ifx\csname natexlab\endcsname\relax\def\natexlab#1{#1}\fi

\bibitem[{{Barker}(2011)}]{Barker2011MNRAS}
{Barker}, A.~J. 2011, \mnras, 414, 1365

\bibitem[{{Christensen} {et~al.}(2020){Christensen}, {Wicht}, \&
  {Dietrich}}]{Christensen2020ApJ}
{Christensen}, U.~R., {Wicht}, J., \& {Dietrich}, W. 2020, \apj, 890, 61

\bibitem[{{Debras} \& {Chabrier}(2019)}]{Debras2019ApJ}
{Debras}, F. \& {Chabrier}, G. 2019, \apj, 872, 100

\bibitem[{{Dewberry} \& {Lai}(2022)}]{Dewberry2022ApJ}
{Dewberry}, J.~W. \& {Lai}, D. 2022, \apj, 925, 124

\bibitem[{{Dewberry} {et~al.}(2021){Dewberry}, {Mankovich}, {Fuller}, {Lai}, \&
  {Xu}}]{Dewberry2021PSJ}
{Dewberry}, J.~W., {Mankovich}, C.~R., {Fuller}, J., {Lai}, D., \& {Xu}, W.
  2021, \psj, 2, 198

\bibitem[{Dintrans {et~al.}(1999)Dintrans, Rieutord, \&
  Valdettaro}]{Dintrans1999}
Dintrans, B., Rieutord, M., \& Valdettaro, L. 1999, Journal Of Fluid Mechanics,
  398, 271

\bibitem[{{Durante} {et~al.}(2020){Durante}, {Parisi}, {Serra}, {Zannoni},
  {Notaro}, {Racioppa}, {Buccino}, {Lari}, {Gomez Casajus}, {Iess}, {Folkner},
  {Tommei}, {Tortora}, \& {Bolton}}]{Durante2020}
{Durante}, D., {Parisi}, M., {Serra}, D., {et~al.} 2020, \grl, 47, e86572

\bibitem[{{Gastine} \& {Wicht}(2021)}]{Gastine2021Icar}
{Gastine}, T. \& {Wicht}, J. 2021, \icarus, 368, 114514

\bibitem[{{Gaulme} {et~al.}(2011){Gaulme}, {Schmider}, {Gay}, {Guillot}, \&
  {Jacob}}]{Gaulme2011AA}
{Gaulme}, P., {Schmider}, F.~X., {Gay}, J., {Guillot}, T., \& {Jacob}, C. 2011,
  \aap, 531, A104

\bibitem[{{Gavrilov} \& {Zharkov}(1977)}]{Gavrilov1977Icar}
{Gavrilov}, S.~V. \& {Zharkov}, V.~N. 1977, \icarus, 32, 443

\bibitem[{Greenspan(1968)}]{Greenspan1968}
Greenspan, H.~P. 1968, {The Theory of Rotating Fluids} (London: Cambridge
  University Press)

\bibitem[{{Guillot} {et~al.}(2004){Guillot}, {Stevenson}, {Hubbard}, \&
  {Saumon}}]{Guillot2004}
{Guillot}, T., {Stevenson}, D.~J., {Hubbard}, W.~B., \& {Saumon}, D. 2004, in
  Jupiter. The Planet, Satellites and Magnetosphere, ed. F.~{Bagenal}, T.~E.
  {Dowling}, \& W.~B. {McKinnon}, Vol.~1, 35--57

\bibitem[{{Idini} \& {Stevenson}(2021)}]{Idini2021PSJ}
{Idini}, B. \& {Stevenson}, D.~J. 2021, \psj, 2, 69

\bibitem[{{Idini} \& {Stevenson}(2022{\natexlab{a}})}]{Idini2022PSJLost}
{Idini}, B. \& {Stevenson}, D.~J. 2022{\natexlab{a}}, \psj, 3, 11

\bibitem[{{Idini} \& {Stevenson}(2022{\natexlab{b}})}]{Idini2022PSJ}
{Idini}, B. \& {Stevenson}, D.~J. 2022{\natexlab{b}}, \psj, 3, 89


\bibitem[{{Lai}(2021)}]{Lai2021PSJ}
{Lai}, D. 2021, \psj, 2, 122

\bibitem[{{Lainey} {et~al.}(2009){Lainey}, {Arlot}, {Karatekin}, \& {van
  Hoolst}}]{Lainey2009}
{Lainey}, V., {Arlot}, J.-E., {Karatekin}, {\"O}., \& {van Hoolst}, T. 2009,
  \nat, 459, 957

\bibitem[{{Lin} \& {Ogilvie}(2017)}]{Lin2017MNRAS}
{Lin}, Y. \& {Ogilvie}, G.~I. 2017, \mnras, 468, 1387

\bibitem[{{Lin} \& {Ogilvie}(2018)}]{Lin2018MNRAS}
{Lin}, Y. \& {Ogilvie}, G.~I. 2018, \mnras, 474, 1644

\bibitem[{{Lin} \& {Ogilvie}(2021)}]{Lin2021ApJ}
{Lin}, Y. \& {Ogilvie}, G.~I. 2021, \apjl, 918, L21

\bibitem[{{Lockitch} \& {Friedman}(1999)}]{Lockitch1999ApJ}
{Lockitch}, K.~H. \& {Friedman}, J.~L. 1999, \apj, 521, 764

\bibitem[{{Militzer} {et~al.}(2022){Militzer}, {Hubbard}, {Wahl}, {Lunine},
  {Galanti}, {Kaspi}, {Miguel}, {Guillot}, {Moore}, {Parisi}, {Connerney},
  {Helled}, {Cao}, {Mankovich}, {Stevenson}, {Park}, {Wong}, {Atreya},
  {Anderson}, \& {Bolton}}]{Militzer2022PSJ}
{Militzer}, B., {Hubbard}, W.~B., {Wahl}, S., {et~al.} 2022, \psj, 3, 185

\bibitem[{{Ogilvie}(2009)}]{Ogilvie2009MNRAS}
{Ogilvie}, G.~I. 2009, \mnras, 396, 794

\bibitem[{{Ogilvie}(2013)}]{Ogilvie2013MNRAS}
{Ogilvie}, G.~I. 2013, \mnras, 429, 613

\bibitem[{{Ogilvie}(2014)}]{Ogilvie2014}
{Ogilvie}, G.~I. 2014, \araa, 52, 171

\bibitem[{{Ogilvie} \& {Lin}(2004)}]{Ogilvie2004ApJ}
{Ogilvie}, G.~I. \& {Lin}, D.~N.~C. 2004, \apj, 610, 477

\bibitem[{{Peale} {et~al.}(1979){Peale}, {Cassen}, \& {Reynolds}}]{Peale1979}
{Peale}, S.~J., {Cassen}, P., \& {Reynolds}, R.~T. 1979, Science, 203, 892

\bibitem[{{Rieutord} {et~al.}(2001){Rieutord}, {Georgeot}, \&
  {Valdettaro}}]{Rieutord2001JFM}
{Rieutord}, M., {Georgeot}, B., \& {Valdettaro}, L. 2001, Journal of Fluid
  Mechanics, 435, 103

\bibitem[{{Stevenson}(2020)}]{Stevenson2020}
{Stevenson}, D.~J. 2020, Annual Review of Earth and Planetary Sciences, 48, 465

\bibitem[{{Stevenson} {et~al.}(2022){Stevenson}, {Bodenheimer}, {Lissauer}, \&
  {D'Angelo}}]{Stevenson2022PSJ}
{Stevenson}, D.~J., {Bodenheimer}, P., {Lissauer}, J.~J., \& {D'Angelo}, G.
  2022, \psj, 3, 74

\bibitem[{Stewartson \& Rickard(1969)}]{Stewartson1969}
Stewartson, K. \& Rickard, J.~A. 1969, Journal of Fluid Mechanics, 35, 759

\bibitem[{{Wahl} {et~al.}(2017){Wahl}, {Hubbard}, {Militzer}, {Guillot},
  {Miguel}, {Movshovitz}, {Kaspi}, {Helled}, {Reese}, {Galanti}, {Levin},
  {Connerney}, \& {Bolton}}]{Wahl2017GeoRL}
{Wahl}, S.~M., {Hubbard}, W.~B., {Militzer}, B., {et~al.} 2017, \grl, 44, 4649

\bibitem[{{Wahl} {et~al.}(2020){Wahl}, {Parisi}, {Folkner}, {Hubbard}, \&
  {Militzer}}]{Wahl2020ApJ}
{Wahl}, S.~M., {Parisi}, M., {Folkner}, W.~M., {Hubbard}, W.~B., \& {Militzer},
  B. 2020, \apj, 891, 42

\bibitem[{{Wei}(2022)}]{Wei2022AA}
{Wei}, X. 2022, \aap, 664, A10

\bibitem[{{Weinberg} {et~al.}(2012){Weinberg}, {Arras}, {Quataert}, \&
  {Burkart}}]{Weinberg2012ApJ}
{Weinberg}, N.~N., {Arras}, P., {Quataert}, E., \& {Burkart}, J. 2012, \apj,
  751, 136

\bibitem[{{Wu}(2005{\natexlab{a}})}]{Wu2005ApJI}
{Wu}, Y. 2005{\natexlab{a}}, \apj, 635, 674

\bibitem[{{Wu}(2005{\natexlab{b}})}]{Wu2005ApJII}
{Wu}, Y. 2005{\natexlab{b}}, \apj, 635, 688

\bibitem[{{Xu} \& {Lai}(2017)}]{Xu2017PhRvD}
{Xu}, W. \& {Lai}, D. 2017, \prd, 96, 083005

\end{thebibliography}

\end{document}